\documentclass[american, aps,12pt ]{revtex4-1} 
\usepackage{color}
\usepackage{times}
\usepackage{subfig}
\usepackage{bm}
\usepackage{graphicx}
\usepackage{graphics}
\DeclareGraphicsExtensions{.pdf,.png,.jpg}
\usepackage{amsbsy}
\usepackage{amsmath}
\usepackage{amsfonts}
\usepackage{amsthm}
\usepackage{float}
\usepackage{amssymb}
\usepackage{textcomp}
\usepackage{mathtools}
\usepackage{hyperref}
\usepackage[font=small,labelfont=bf,
justification= centerlast,
format=hang]{caption}
\newtheorem{innercustomthm}{Theorem}
\newenvironment{customthm}[1] 
  {\renewcommand\theinnercustomthm{#1}\innercustomthm}
  {\endinnercustomthm}

\usepackage[subnum]{cases}
\usepackage{enumitem}
\usepackage[font=small,labelfont=bf,
justification= centerlast,
format=hang]{caption}

\definecolor{purple(html/css)}{rgb}{0.5, 0.0, 0.5}
\newcommand{\ket}[1]{| #1 \rangle}
\newcommand{\bra}[1]{\langle #1 |}

\newcommand{\ketbra}[2]{| #1 \rangle \langle #2 |}

\newcommand \inner[1]{ \left<#1\right>}
\usepackage{enumerate}
\usepackage{braket}
\usepackage{adjustbox}
\usepackage{footnote}
\usepackage[dvipsnames]{xcolor}
\usepackage{tipa}
\usepackage{textcomp}
\usepackage{fontenc}
\usepackage{mathrsfs}
\usepackage{color}
\usepackage{colortbl}
\usepackage{graphics,graphicx}
\usepackage{amsmath}
\usepackage{bbm}

\usepackage{times}

\newtheorem{eg}{Example}
\newtheorem{proposition}{Proposition}

\newtheorem{theorem}{Theorem}
\newtheorem{dfn}{Definition}
\newenvironment{prof}{\noindent \textbf{Proof: }\ignorespaces}{\hspace*{\fill}$\Box$\medskip}
\DeclareOldFontCommand{\rm}{\normalfont\rmfamily}{\mathrm}

\begin{document}

\title{Duality between quantum channels and super-channels is basis-dependent}

\author{Sohail\(^1\), Sahil\(^{2,3}\), Ritabrata Sengupta\(^4\), and Ujjwal Sen\(^1\)}
\affiliation{\(^1\)Harish-Chandra Research Institute,  A CI of Homi Bhabha National Institute, Chhatnag Road, Jhunsi, Allahabad 211  019, India\\
\(^2\)Optics and Quantum Information Group, The Institute of Mathematical Sciences, CIT Campus, Taramani, Chennai 600 113, India\\
\(^3\)Homi Bhabha National Institute, Training School Complex, Anushakti Nagar, Mumbai 400 085, India\\
\(^4\)Department of Mathematical Sciences, Indian Institute of Science, Education \& Research (IISER) Berhampur, Transit Campus, Govt. ITI, NH 59, Berhampur 760 010, Ganjam, Odisha, India}

\email{$^1$sohail@hri.res.in}
\email{$^2$sahilmd@imsc.res.in}
\email{$^4$rb@iiserbpr.ac.in}\
\email{$^1$ujjwal@hri.res.in}

\begin{abstract}
The complete positivity vs positivity correspondence in the Choi-Jamio{\l}kowski-Kraus-Sudarshan quantum channel-state isomorphism depends on the choice of basis. Instead of the ``canonical'' basis, if we use, e.g., the Pauli spin matrices along with the identity as the basis for the space of bounded operators on the two-dimensional complex Hilbert space, this correspondence breaks down. A sufficient condition on the basis for  validity of this correspondence is provided in the work of Paulsen and Shultz~\cite{Paulsen}, which was  later proven to be necessary by Kye~\cite{Kye}. A correspondence is also present between the space of super-maps and the tensor product of the spaces of the inputs and outputs of the same. In particular, a super-map is completely CP-preserving if and only if its Choi-type representation is completely positive (CP). This correspondence also depends on a specific choice of basis. In this work, we find the necessary and sufficient condition on a basis such that this correspondence  holds true. 
\end{abstract}
\maketitle
\section{introduction}
 In quantum mechanics, a physical system is represented by a complex separable Hilbert space. 
The state of the physical system is described by a positive semi-definite trace-class operator with unit trace, and the physical quantities that we are interested to measure are conveniently described by hermitian operators on the same Hilbert space. If the system is isolated from its environment, the evolution of the system is described by a unitary operator. However, in the real world, no physical system is completely isolated from its environment. To treat such a situation, the system and environment are evolved under a joint unitary, 
and the environment  is traced out from the system-environment final state in order to obtain the evolution of the system's state. This approach gives rise to the quantum channel approach to general evolution of a quantum system. Any physical process that a quantum system goes through is described by a quantum channel, which is nothing but a completely positive  trace-preserving map. The notion of positivity and complete positivity are 
crucially
important concepts in quantum information  as well as in operator algebra. Given two Hilbert spaces, $H_1$ and $H_2$, a map $\Phi : B(H_1) \rightarrow {B}(H_2)$ is called positive if it maps positive elements of $B(H_1)$ to the positive elements of ${B}(H_2)$. Now using $\Phi$, one can construct a map $id_{k} \otimes\Phi:M_k(\mathbb{C}) \otimes B(H_1) \rightarrow M_k(\mathbb{C}) \otimes B(H_2)$, where $M_k(\mathbb{C})$ is the space of $k \times k$ matrices whose entries are complex numbers. The map $\Phi$ is called \(k\)-positive if $id_{k} \otimes\Phi$ maps positive elements of $M_k(\mathbb{C}) \otimes B(H_1)$ to positive elements of $M_k(\mathbb{C}) \otimes B(H_2)$. If $\Phi$ is \(k\)-positive for all $k \in \mathbb{N}$, then we call it a completely positive map. It is clear from the definition that a complete positive map is a positive map, however the converse is not always true. A standard  example of a positive but not completely positive map is the transposition map on matrices. Using this property of transposition, one can detect whether certain bipartite quantum states has entanglement or not \cite{HORODECKI19961,PhysRevLett.77.1413}. For a given map, to know whether it is completely positive we employ the Choi theorem on completely positive maps, which says that a map is completely positive if and only if its Choi matrix is positive. The definition of the Choi matrix has intrinsic connections with a chosen basis, 
the ``canonical'' basis, which is generally taken as a maximally entangled pure state. 
Paulsen and Shultz~\cite{Paulsen}, and later Kye~\cite{Kye}, provided the necessary and sufficient condition on a chosen basis such that the Choi theorem holds true.
\par Quantum channels are one of the important entities in quantum mechanics as well as in quantum information theory. Quantum channels can be viewed as a quantum state with one dimensional input~\cite{Chiribella}. In particular a quantum state with a pure state decomposition $\{ p_i, \ket{\psi_{i}}\}$ can be thought as a quantum channel with karus operators $\sqrt{p_i}\ket{\psi_{i}}$. Quantum states are actually a preparation procedure from the operational point of view. Hence quantum channels can be considered as the fundamental entity in quantum mechanics describing both static and dynamic processes. Quantum channels may undergo transformation due to various reasons. The dynamics of quantum channels are described by super-maps called quantum super-channels. When a quantum super-channel acts on one part of an arbitrary bipartite quantum channel, we expect the resulting object to be a quantum channel again. So a super-map has to satisfy two conditions in order to become a quantum super-channel: (i) it has to be completely complete positivity-preserving (completely CP-preserving) and (ii) it has to preserve the trace-preserving nature of maps (TP-preserving). A Choi-type theorem holds here as well. It states that a super-map is completely CP-preserving if and only if its Choi-type representation is completely positive. We observe that the 
validity
of this theorem also depends on the choice of basis. 
We give the necessary and sufficient condition on a chosen basis such that the Choi-type theorem between super-maps and maps holds true. Quantum super-channels appears at various contexts in quantum information theory such as quantum channel discrimination problem, entropy of a quantum channel~\cite{Gour_entropy_channel}, entanglement-breaking super-channels~\cite{Entanglement-breakin-superchannels}, measurement of a quantum channel~\cite{Measurement_on_channels} etc. In Channel discrimination problem Alice is given an arbitrarily chosen channel from a set of two channels $\{ \Phi_1, \Phi_2 \}$ which she needs to distinguish. Now if Alice apply some super-channel to her quantum channel then the distinguishability should not increase. If $D(\Phi_1 || \Phi_2)$ is a measure of distinguishability the under the action of any super-channel $\Theta$ it should be non-increasing, i.e., the the monotonicity relation $D(\Theta(\Phi_1) || \Theta(\Phi_2)) \leq D(\Phi_1 || \Phi_2)$ should hold always. While defining entropy of a quantum channel the entropy, it is defined in such a way that under random unitary super-channels the entropy of a quantum channel always increases. A bipartite quantum channel can be classified as separable and nonseparable quantum channels. Separable quantum channels are the one that can not produce entangled state while acted on a separable state. The nonseparable quantum channels are the one that can produce entangled state from a separable state. Let Alice and Bob shares a bipartite quantum channel. Alice and Bob can manipulate their part of the shared channel locally which is described by local super-channel. If their shared quantum channel is separable then any local super-channel applied by them can not turn it into a non-separable channel. However if the shared quantum channel is non-separable then it may be possible that after the application of a local super-channel the shared channel becomes separable. If a quantum super-channel when acted upon locally on any bipartite quantum channel make it a separable quantum channel, then it is known as a entanglement-breaking super-channel. Study of entanglement-breaking super-channel is important in quantum information theory. The concept of measurement of a quantum channel is formulated using quantum super-channels. In particular a measurement on a quantum channel is a collection of CPTNI( completely positive trace non-increasing)- preserving super-maps such that their addition is a quantum super-channel. While working with quantum super-channels it is usually convenient to work with the Choi matrix of it, which is nothing but the Choi matrix of the Choi-type representation of the super-channel. Notice that two steps are involved in going from quantum super channels to the Choi matrix representation of it: first we need the Choi-type representation of the super channel then we need the Choi matrix of the Choi-type representation. In the existing literature the Choi matrix for super map is constructed using the so called ``canonical basis" in each of the two steps. In this work we show that if a chosen basis satisfies the necessary and sufficient condition we provide then it is good enough to be used for the study of quantum super-channels. The work of Paulsen,Shultz and Kye allows flexibility in the choice of basis while going from Choi-type representation to the Choi-matrix representation. Our work allows flexibility in the choice basis while going from super-super channel to its Choi-type representation, thus providing a full flexibility in the choice of basis while going from a super channel to its Choi matrix representation. 
\par The paper is organised as follows: In Section \ref{s2} we give the basic definitions, viz. the basics ideas of Choi-Jamio{\l}kowski isomorphism and the dependence of the CP vs positivity correspondence on the choice of basis. In Section \ref{s3}, we discuss the necessary ideas about super-maps and present our main results. In Section \ref{s4}, we conclude our paper.

\section{The CJKS Isomorphism} \label{s2}

Let \(H_1\) and \(H_2\) be two finite-dimensional Hilbert spaces and $B(H_1)$ and $B(H_2)$ represent the space of bounded linear operators on them respectively,
and let $\Phi :B(H_1)\rightarrow B(H_2)$ be a linear map, where 
\(H_1 \equiv \mathbb{C}^n\) and \(H_2 \equiv \mathbb{C}^m\).
Let $ \{e_{ij}\}$, $i,j=1,2,\ldots,n$ be a complete set of matrix units for $B(H_1)$ and $ \{f_{ij}\}$, $i,j=1,2,\ldots,m$ be a complete set of matrix units for $B(H_2)$   Then the CJKS  matrix \cite{book,article,CHOI1975285,kraus1983states,Sudarshan1985} for $\phi$ is defined as the operator
\begin{eqnarray}
        C_\Phi =  \sum_{{i,j}=1}^{n} e_{ij} \otimes \Phi(e_{ij}) \in B(H_1)\otimes B(H_2). \label{CJKSdefn}
\end{eqnarray}                   
 It is clear that the map $J: L(B(H_1),B(H_2)) \rightarrow B(H_1)\otimes B(H_2)$ defined by  $J(\Phi)= C_\Phi $ is linear and injective. Now $X \in B(H_1)\otimes B(H_2)$ can be written as 
 \begin{eqnarray}
     X&&=\sum_{i,j,k,l} X_{ijkl}  e_{ij}\otimes f_{kl} \\
     &&= \sum_{i,j} e_{ij}\otimes \sum_{k,l} X_{ijkl}f_{kl}.
 \end{eqnarray}
 Now  
 \begin{eqnarray}
     \Gamma_X(e_{ij})=\sum_{k,l} X_{ijkl}f_{kl}
 \end{eqnarray}
  defines a map $\Gamma_X :B(H_1)\rightarrow B(H_2)$ by linear extension. So we can write
 \begin{eqnarray}
     X=\sum_{{i,j}=1}^{n} e_{ij} \otimes \Gamma_{X}(e_{ij})
 \end{eqnarray}
  which implies that the map $J$ is surjective as well. This map $J$ is known as Choi-Jamio{\l}kowski-Kraus-Sudarshan (CJKS) isomorphism. The action of $\Gamma_{X}$ on an element $A\in B(H_1)$ can be written as the following:
  \begin{eqnarray}
      \Gamma_{X}(A)=tr_{1}[(A^t \otimes \mathbb{I})X],
  \end{eqnarray}
  where $t$ represent transposition with respect to the usual canonical basis of $H_1$ and $\mathbb{I}$ represents the identity operator on $H_2$. Although the CJKS isomorphism is a vector space isomorphism, it can be thought as an algebra isomorphism between the convolution algebra $L(B(H_1),B(H_2))$ and the matrix algebra $B(H_1)\otimes B(H_2)$ as shown in~\cite{Sohail_2022}.
  
  This isomorphism helps us to formulate the concept of a ``channel-state duality". Below we state the Choi theorem on completely positive maps, which is necessary to understand the concept of ``channel-state duality".\\
 

\begin{customthm}{A}[Choi-Jamio{\l}kowski isomorphism] \cite{book,article,CHOI1975285,kraus1983states,Sudarshan1985}.
The CJKS matrix $C_\Phi = $ $\sum_{{i,j}=1}^{n} $ $e_{ij} \otimes \Phi(e_{ij})$ $\in $ $B(H_1)\otimes B(H_2) $ is positive if and only if the map $\Phi :B(H_1)\rightarrow B(H_2)$ is completely positive.
\end{customthm}

We can view completely positive trace-preserving linear maps acting on quantum states as a quantum state with maximally mixed marginal in a higher-dimensional Hilbert space thanks to the CJKS isomorphism and the Choi theorem on completely positive maps. If we consider quantum states which are density matrices on an $n$-dimensional Hilbert space, then the quantum channels acting on them can be identified with a quantum state with maximally mixed marginals on an  $n^2$-dimensional Hilbert space. From now we will refer the Choi theorem on complete positivity as ``CP vs positivity" correspondence.
\subsection{Dependence on basis of the CP vs positivity correspondence}
It is clear from the definition (\ref{CJKSdefn}) that the CJKS matrix is defined with respect to a particular basis, namely the ``canonical" basis $\{e_{ij}\}$. Instead, if we use a different set of basis to define the CJKS matrix $C_{\Phi}$ then also the spaces $L(B(H_1), B(H_2))$ and $B(H_1)\otimes B(H_2)$ are isomorphic under the identification $\Phi \leftrightarrow C_{\Phi}$, but the CP vs positivity correspondence may not hold true. In the following example we demonstrate this fact.
\begin{eg} Let us choose the Hilbert space $H_1$ and $H_2$ to be two-dimensional and let us choose the basis for $B(H_1)$ to be the Pauli spin matrices along with identity matrix, i.e. $\{  \mathbb{I}, \sigma_x, \sigma_y, \sigma_z\}$. Now instead of the canonical basis $\{e_{ij}\}$, if we use the Pauli basis in the Choi-Jamio{\l}kowski isomorphism we have $\Phi \leftrightarrow \Tilde{C_{\Phi}}:=\mathbb{I} \otimes \Phi(\mathbb{I}) +  \sigma_x \otimes \Phi( \sigma_x) + \sigma_y \otimes \Phi( \sigma_y)+\sigma_z \otimes \Phi( \sigma_z)$ as a vector space isomorphism. Now choosing $\Phi=id$, the identity map,  which is obviously completely positive, we have $\Tilde{C_{id}}=\mathbb{I} \otimes \mathbb{I} +  \sigma_x \otimes \sigma_x + \sigma_y \otimes \sigma_y+\sigma_z \otimes \sigma_z$ which have eigenvalues $\{ 2,2,2,-2\}$. So this example shows that CP vs positivity correspondence does not hold for an arbitrarily chosen basis.
\end{eg}
\begin{eg} 
Let $\{\ket{i}\}$ and $\{\ket{\lambda_j}\}$ be two different orthonormal basis for the Hilbert space $H_1$. We can construct an orthonormal basis for $B(H_1)$ as $E_{ij}:=\ketbra{i}{\lambda_j}$. With this basis we have $\Phi \leftrightarrow \Tilde{C_{\Phi}}:=\sum_{ij} E_{ij} \otimes \Phi(E_{ij})$ as a vector space isomorphism. Now taking $\Phi=id$ which is completely positive, we have $\Tilde{C}_{id}=\sum_{ij} E_{ij} \otimes E_{ij}= (\sum_{i} \ket{i} \otimes \ket{i}) (\sum_{j} \bra{\lambda_{j}} \otimes \bra{\lambda_{j}})$ which is clearly not positive.
\end{eg}
\par Now it is natural to ask the question that what is the necessary and sufficient condition on a chosen basis such that the CP vs positivity correspondence holds? The work of Paulsen and Shultz \cite{Paulsen} and later the work of Kye \cite{Kye} provides the answer. In the following paragraph we provide a brief discussion of their main results.
\par Let $\{h_{ij}\}$ be an arbitrary basis for $B(H_1)$ and let $W$ be the map that maps $\{e_{ij}\}$ to $\{h_{ij}\}$ i.e., $W(e_{ij})=h_{ij}$. Let us define $M:=WW^t :B(H_1) \rightarrow B(H_1)$ where $W^t$ is the map whose matrix representation is the transposition of the matrix representation of the map $W$. The following theorems gives the necessary and sufficient condition on the basis $\{h_{ij}\}$ so that the CP vs positivity correspondence holds true.
\begin{dfn}
    A map $\Phi : B(H_1) \rightarrow B(H_2)$ is called complete order isomorphism if both $\Phi$ and $\Phi^{-1}$ are completely positive.
\end{dfn}

\begin{theorem}\cite{Kye,Paulsen}
    The CP vs positivity correspondence holds for the basis $\{h_{ij}\}$ of $B(H_1)$ if and only if $M$ is a complete order isomorphism.
\end{theorem}
\noindent The following theorem tells us how a complete order isomorphism looks like.
\begin{theorem} \cite{Paulsen,Kye}
   A map $\Phi: B(H_1) \rightarrow B(H_2)$ is a complete order isomorphism if and only if there exists invertible $K: H_1 \rightarrow H_2$ such that $\Phi(X)= K X K^{\dagger}$ \hspace{0.2cm} $\forall X \in B(H_1).$ \label{KPCOI}
\end{theorem}
{With the help of the above two theorems the structure for the basis $\{ h_{ij} \}$ can be given by the following theorem.
\begin{theorem} \cite{Kye}
    For a basis $\{ h_{ij} \}$ the CP vs positivity correspondence holds true if and only if there exists a basis $\{ \zeta_{i}\}$ for $H_1$ such that $h_{ij}=\ketbra{\zeta_{i}}{\zeta_{j}}$. \label{cannonicalness}
\end{theorem}}
\section{Space of Super-maps}\label{s3}
We consider the Hilbert spaces $H_1$, $H_2$, $H_3$ and $H_4$ to be finite dimensional. Let $B(H_i)$ represents the space of bounded linear operators on the Hilbert space $H_i$, for $i=1,2,3,4$. We represent by $L(B(H_i), B(H_j))$ the space of linear operators from $B(H_i)$ to $B(H_j)$ and by $L[L(B(H_1),B(H_2)),L(B(H_3),B(H_4))]$ we represent the space of linear operators from $L(B(H_1),B(H_2))$ to $L(B(H_3),B(H_4))$. A linear map  $\Theta :L(B(H_1),B(H_2)) \rightarrow L(B(H_3),B(H_4))$ is called a super-map.
\\
\subsection{Choi-type theorem for Super-maps}
Here we choose canonical basis for the spaces $B(H_{1})$, $B(H_{2})$,$B(H_{3})$,$B(H_{4})$ and denote them by $e^{H_n}_{ij}$, where $n=1,2,3,4$. In the bra-ket notation $e^{H_n}_{ij}=\ketbra{i_{H_n}}{j_{H_n}}$ where $\{\ket{i_{H_n}}\}$ is an orthonormal basis for $H_n$. The orthonormal basis for $L(B(H_1), B(H_2))$ is the following\cite{Gour}
\begin{eqnarray}
    \mathcal{E}^{H_1 \rightarrow H_2}_{ijkl}(A)
    &&= tr({e_{ij}^{H_1}}^{\dagger}A)e^{H_2}_{kl}, \label{canonical_basis_super-map}
\end{eqnarray}
where ``$\dagger$" denotes the hermitian conjugate.
Let us denote by $\mathbb{CP}[1 \rightarrow 2]$ and $\mathbb{CP}[3 \rightarrow 4]$ the convex cone of completely positive maps of $L(B(H_1),B(H_2))$ and $L(B(H_3),B(H_4))$ respectively. Now we provide necessary definitions and theorems in order to proceed towards our main results.
\begin{dfn}
    A super-map $\Theta :L(B(H_1),B(H_2)) \rightarrow L(B(H_3),B(H_4))$ is called CP-preserving if it maps CP maps in $\mathbb{CP}[1 \rightarrow 2]$ to the CP maps in $\mathbb{CP}[3 \rightarrow 4]$, i.e., $\Theta(\mathbb{CP}[1 \rightarrow 2]) \subseteq \mathbb{CP}[3 \rightarrow 4]$.
\end{dfn}
\begin{dfn}
    A super-map $\Theta :L(B(H_1),B(H_2)) \rightarrow L(B(H_3),B(H_4))$ is called completely CP preserving if the map $id \otimes \Theta : L(B(K_1),B(K_2)) \otimes L(B(H_1),B(H_2)) \rightarrow L(B(K_1),B(K_2)) \otimes L(B(H_3),B(H_4))$ is CP-preserving for for all choices of Hilbert spaces $K_1$ and $K_2$.
\end{dfn}
\begin{dfn}
    A super-map $\Theta :L(B(H_1),B(H_2)) \rightarrow L(B(H_3),B(H_4))$ is called complete order isomorphism if both $\Theta$ and $\Theta^{-1}$ are completely CP-preserving.
\end{dfn}
The spaces $L\left[L(B(H_1),B(H_2)), L(B(H_3),B(H_4))\right]$ and $L(B(H_1),B(H_2)) \otimes L(B(H_3),B(H_4))$ are isomorphic as vector spaces under the following identification
\begin{eqnarray}
    \Theta \leftrightarrow \sum_{ijkl}  \mathcal{E}^{H_1 \rightarrow H_2}_{ijkl} \otimes \Theta( \mathcal{E}^{H_1 \rightarrow H_2}_{ijkl}).
\end{eqnarray}
It should be noted that the object $\Lambda_{\Theta}=:\sum_{ijkl}  \mathcal{E}^{H_1 \rightarrow H_2}_{ijkl} \otimes \Theta( \mathcal{E}^{H_1 \rightarrow H_2}_{ijkl}) \in L(B(H_1),B(H_2)) \otimes L(B(H_3),B(H_4)) $ has the mathematical structure similar to Choi matrices. The theorem below supports this similarity further.
\begin{theorem}
    \cite{Chiribella,Gour} The super-map $\Theta :L(B(H_1),B(H_2)) \rightarrow L(B(H_3),B(H_4))$ is completely CP-preserving if and only if $\Lambda_{\Theta}=:\sum_{ijkl}  \mathcal{E}^{H_1 \rightarrow H_2}_{ijkl} \otimes \Theta( \mathcal{E}^{H_1 \rightarrow H_2}_{ijkl}) $ is CP. \label{Choi-type_theorem}
\end{theorem}
We will refer to $\Lambda_{\Theta}$ as the Choi-type representation of the super-map $\Theta$ in this paper.
\\
Now if we use a different basis $\{\mathcal{F}_{\beta}\}$ for the space $L(B(H_1),B(H_2))$, then also the spaces $L\left[L(B(H_1),B(H_2)), L(B(H_3),B(H_4))\right]$ and $L(B(H_1),B(H_2)) \otimes L(B(H_3),B(H_4))$ are isomorphic as vector spaces under the following identification
\begin{eqnarray}
    \Theta \leftrightarrow \sum_{\beta}  \mathcal{F}^{H_1 \rightarrow H_2}_{\beta} \otimes \Theta( \mathcal{F}^{H_1 \rightarrow H_2}_{\beta}).
\end{eqnarray}
We are now interested to know whether the following correspondence is true for the basis $\{ \mathcal{F}^{H_1 \rightarrow H_2}_{\beta}\}$:
\begin{center}
   \emph{ The super-map $\Theta :L(B(H_1),B(H_2)) \rightarrow L(B(H_3),B(H_4))$ is completely\\ CP-preserving if and only if $\Lambda^{\mathcal{F}}_{\Theta}=:\sum_{\beta}  \mathcal{F}^{H_1 \rightarrow H_2}_{\beta} \otimes \Theta( \mathcal{F}^{H_1 \rightarrow H_2}_{\beta}) $ is CP.}
\end{center}
From now on we will use ``CCPP vs CP correspondence" to mean ``Completely CP-preserving if and only if CP".

We have an explicit example that shows that this CCPP vs CP correspondence may not hold for an arbitrarily chosen basis. In particular the following example shows that even if we choose a basis whose mathematical form is similar to basis defined in Eq.~(\ref{canonical_basis_super-map}) except the ``canonical" basis $\{e_{ij}\}$ replaced by some arbitrary basis then also the CCPP vs CP correspondence may not hold true.
\begin{eg}
For a given pair of bases $\{b^{1}_{i}\}$ and $\{b^{2}_{j}\}$ for $B(H_1)$ and $B(H_2)$ respectively we can define basis similar to Eq.~(\ref{canonical_basis_super-map}) for the space $L(B(H_1),B(H_2))$ as follows,
\begin{eqnarray}
    \mathcal{F}^{H_1 \rightarrow H_2}_{ij}(A)
    &&= tr(b^{1\dagger}_{i}A)b^{2}_{j} \hspace{1cm} \forall A \in B(H_1). \label{cannonical_basis_map}
\end{eqnarray}
Now, we consider both $H_1$ and $H_2$ to be two-dimensional and we choose the Pauli matrices including  the identity matrix as a basis for both $B(H_1)$ and $B(H_2)$. Let $\sigma_0 = \mathbb{I}$, $\sigma_1= \sigma_x$, $\sigma_2 = \sigma_y$ and $\sigma_3= \sigma_z$. So a basis for $L(B(H_1), B(H_2))$ is given by $ \mathcal{F}^{H_1 \rightarrow H_2}_{ij}(A)= tr(\sigma_{i}^{\dagger}A)\sigma_{j}$.
Now choosing $\Theta = {id}$, which is completely CP-preserving map, we have $\Lambda^{\mathcal{F}}_{{id}}=\sum_{ij}  \mathcal{F}^{H_1 \rightarrow H_2}_{ij} \otimes \mathcal{F}^{H_1 \rightarrow H_2}_{ij}$. The Choi matrix for $\Lambda^{\mathcal{F}}_{{id}}$ turns out to be $C_{\Lambda^{\mathcal{F}}_{{id}}}=\sum_i \Bar{\sigma_{i}} \otimes \Bar{\sigma_{i}} \otimes \sum_j \sigma_{j} \otimes \sigma_{j} $ (here $\Bar{\sigma_{i}}$ represents complex conjugate of $\sigma_{i}$ ) which have six negative eigen values implying that $\Lambda^{\mathcal{F}}_{{id}}$ is not completely positive. So for this particular choice of basis $\{\mathcal{F}^{H_1 \rightarrow H_2}_{ij}\}$ the CCPP vs CP correspondence does not hold. {While it may be tempting to believe from Theorem~(\ref{cannonicalness}) that the CCPP vs CP correspondence will hold true for the basis given in Eq~(\ref{cannonical_basis_map}), this specific example demonstrates that it does not.}
\end{eg}
\begin{eg} 
Let $\{e_{ij}\}$ be the usual ``canonical" basis for $B(H_1)$ and $\{E_{ij}\}$ be the basis for $B(H_2)$ as discussed in Example~2. A basis similar to Eq.~(\ref{canonical_basis_super-map}) for the space $L(B(H_1),B(H_2))$ cab be chosen as follows,
\begin{eqnarray}
    \Tilde{\mathcal{F}}^{H_1 \rightarrow H_2}_{ijkl}(A)
    &&= tr(e_{ij}^{\dagger}A)E_{kl} \hspace{1cm} \forall A \in B(H_1). 
\end{eqnarray}
Now choosing $\Theta=id$, have have $\Lambda^{\Tilde{\mathcal{F}}}_{{id}}=\sum_{ijkl}  \Tilde{\mathcal{F}}^{H_1 \rightarrow H_2}_{ijkl} \otimes \Tilde{\mathcal{F}}^{H_1 \rightarrow H_2}_{ijkl}$, whose Choi matrix is $C_{\Lambda^{\Tilde{\mathcal{F}}}_{{id}}}= (\sum_{ij} e_{ij} \otimes e_{ij}) \otimes (\sum_{kl} E_{kl} \otimes E_{kl})$. As $\sum_{kl} E_{kl} \otimes E_{kl}$ is not hermitian, the Choi matrix $C_{\Lambda^{\Tilde{\mathcal{F}}}_{{id}}}$ is not semi-definite. So the CCPP vs CP correspondence does not hold here.
\end{eg}
\par We now ask the question: what is the necessary and sufficient condition on the choice of basis such that the CCPP vs CP correspondence holds? In a broader sense we are interested to know that what happens to the CCPP vs CP correspondence if we replace the map $\sum_{ijkl}  \mathcal{E}^{H_1 \rightarrow H_2}_{ijkl} \otimes \mathcal{E}^{H_1 \rightarrow H_2}_{ijkl} $ by an arbitrary map $\mathcal{G} \in L(B(H_1),B(H_2)) \otimes L(B(H_1),B(H_2))$.
\\
Before proceeding to our main results, we state and prove some elementary results.
\\
\\
We denote by $\mathbb{CCPP}$ the convex cone of completely CP-preserving maps. Following the work of \cite{GIRARD,Kye} we define a bi-linear pairing $\inner{\Theta_1, \Theta_2}$ on the space $L\left[L(B(H_1),B(H_2)), L(B(H_3),B(H_4))\right]$ by 
\begin{eqnarray}
    \inner{\Theta_1, \Theta_2}:=\inner{\Lambda_{\Theta_1},\Lambda_{\Theta_2}}= \inner{C_{\Lambda_{\Theta_1}},C_{\Lambda_{\Theta_2}}} =Tr(C^{t}_{\Lambda_{\Theta_1}}C_{\Lambda_{\Theta_2}}), \label{bilinear}  
\end{eqnarray}
where $\Lambda_{\Theta}$ is the Choi-type representation of $\Theta$ and $C_{\Lambda_{\Theta}}=\sum_{ijkl}(e_{ij}^{H_{1}} \otimes e_{kl}^{H_{3}}) \otimes \Lambda_{\Theta}(e_{ij}^{H_{1}} \otimes e_{kl}^{H_{3}})$ is the Choi matrix of $\Lambda_{\Theta}$.\\
The adjoint of a map $\Theta$ is defined with respect to the above defined bi-linear pairing by the relation
\begin{eqnarray}
    \inner{\Theta^{*}(\Psi), \Phi}= \inner{\Psi, \Theta(\Phi)}.
\end{eqnarray}
It can easily be verified that $(\Theta_{1}\circ\Theta_{2})^{*}=\Theta^{*}_{2} \circ \Theta^{*}_{1}$.
\\
We define the dual cone of $\mathbb{CCPP}$ by $\mathbb{CCPP}^{\circ}:=\{ \Theta : \inner{\Theta, \Omega} \geq 0, \hspace{0.2cm} \forall \Omega \in \mathbb{CCPP} \}$.
\\
\begin{proposition}
The cone $\mathbb{CCPP}$ is self dual, i.e., $\mathbb{CCPP}^{\circ}=\mathbb{CCPP}$  \label{prop1}
\end{proposition}
\begin{prof}
    Let $\Theta \in \mathbb{CCPP}$, Then from Eq. (\ref{bilinear}) we have $\inner{\Theta, \Omega} \geq 0 $ $ \forall \Omega \in \mathbb{CCPP}$, which implies that $\Theta \in \mathbb{CCPP}^{\circ}$. So we have the inclusion $\mathbb{CCPP} \subseteq \mathbb{CCPP}^{\circ}$. Now let $\Theta \in \mathbb{CCPP}^{\circ}$. So $\inner{\Theta, \Omega} \geq 0$ $ \forall \Omega \in \mathbb{CCPP}$, which implies that $Tr(C^{t}_{\Lambda_{\Theta}}C_{\Lambda_{\Omega}}) \geq 0$ $ \forall \Omega \in \mathbb{CCPP}$ which implies $C_{\Lambda_{\Theta}} \geq 0$ which implies $\Lambda_{\Theta} \in \mathbb{CP}$ which implies $\Theta \in \mathbb{CCPP}$. So we have $\mathbb{CCPP}^{\circ} \subseteq \mathbb{CCPP}$. Finally we have $\mathbb{CCPP}=\mathbb{CCPP}^{\circ}$.
\end{prof}
\\
Next two propositions directly follows from the definition (\ref{bilinear}) of the bi-linear pairing.
\begin{proposition}
  The bi-linear pairing defined above is symmetric, i.e., $\inner{\Theta_1, \Theta_2}=\inner{\Theta_2, \Theta_1}$. \label{prop2}
\end{proposition}
\begin{proposition}
    For super-maps $\Theta_{1}$, $\Delta_{1}$ and $\Theta_{2}$, $\Delta_{2}$, we have
    \begin{eqnarray}
    \nonumber
        \inner{\Theta_{1} \otimes \Theta_{2},\Delta_{1} \otimes \Delta_{2}} = \inner{\Theta_{1},\Delta_{1}} \inner{\Theta_{2},\Delta_{2}} 
    \end{eqnarray} \label{prop3}
\end{proposition}
\par For a super-map we have $\Lambda_{\Theta}=\sum_{ijkl}  \mathcal{E}^{H_1 \rightarrow H_2}_{ijkl} \otimes \Theta( \mathcal{E}^{H_1 \rightarrow H_2}_{ijkl})$. The action of $\Theta$ on the basis elements is given by
\begin{eqnarray}
     \Theta( \mathcal{E}^{H_1 \rightarrow H_2}_{ijkl})=\sum_{pqrs} a_{ijklpqrs} \mathcal{E}^{H_3 \rightarrow H_4}_{pqrs}.
\end{eqnarray}
Using the above expression we can rewrite $\Lambda_{\Theta}$ as the following,
\begin{eqnarray}
    \Lambda_{\Theta}=\sum_{ijkl} \sum_{pqrs} a_{ijklpqrs}  \mathcal{E}^{H_1 \rightarrow H_2}_{ijkl} \otimes  \mathcal{E}^{H_3 \rightarrow H_4}_{pqrs}. \label{expansion1}
\end{eqnarray}
We now define a map $\Tilde{\Theta}:L(B(H_3),B(H_4)) \rightarrow L(B(H_1),B(H_2))$ by the following
\begin{eqnarray}
     \Tilde{\Theta}( \mathcal{E}^{H_3 \rightarrow H_4}_{pqrs})=\sum_{ijkl} a_{ijklpqrs} \mathcal{E}^{H_1 \rightarrow H_2}_{ijkl}. \label{defn tilda}
\end{eqnarray}
Using Eq.~(\ref{defn tilda}) in Eq.~(\ref{expansion1}) we have 
\begin{eqnarray}
    \Lambda_{\Theta}=\sum_{pqrs} \Tilde{\Theta}( \mathcal{E}^{H_3 \rightarrow H_4}_{pqrs}) \otimes  \mathcal{E}^{H_3 \rightarrow H_4}_{pqrs}. \label{tilde-choi-type}
\end{eqnarray}
\begin{proposition}
  The adjoint of the map $\Theta$ is equal to the above defined map $\Tilde{\Theta}$ , i.e.,  $\Tilde{\Theta}=\Theta^*$. \label{tilda=star}
\end{proposition}
\begin{prof}
From the definition of the adjoint of a super-map we have
    \begin{eqnarray}
    \nonumber
        \inner{\Theta^{*}( \mathcal{E}^{H_3 \rightarrow H_4}_{pqrs}), \Phi} &&= \inner{\mathcal{E}^{H_3 \rightarrow H_4}_{pqrs}, \Theta(\Phi)}\\
    \nonumber    
        =&& \sum_{tuvw} \Phi_{tuvw} \inner{\mathcal{E}^{H_3 \rightarrow H_4}_{pqrs}, \Theta(\mathcal{E}^{H_1 \rightarrow H_2}_{tuvw})}\\
    \nonumber    
        =&& \sum_{tuvw}\sum_{ijkl} \Phi_{tuvw} a_{tuvwijkl} \inner{\mathcal{E}^{H_3 \rightarrow H_4}_{pqrs}, \mathcal{E}^{H_3 \rightarrow H_4}_{ijkl}}\\
    \nonumber    
         =&& \sum_{tuvw}\sum_{ijkl} \Phi_{tuvw} a_{tuvwijkl} \delta_{ip}\delta_{jq}\delta_{kr}\delta_{ls}\\   
         =&& \sum_{tuvw} \Phi_{tuvw} a_{tuvwpqrs}. \label{tildestar1}
    \end{eqnarray}

Now,
   \begin{eqnarray}
   \nonumber
       \inner{ \Tilde{\Theta}( \mathcal{E}^{H_3 \rightarrow H_4}_{pqrs}), \Phi} &&=  \sum_{tuvw}\sum_{ijkl} \Phi_{tuvw} a_{ijklpqrs} \inner{\mathcal{E}^{H_1 \rightarrow H_2}_{ijkl},\mathcal{E}^{H_1 \rightarrow H_2}_{tuvw}}\\
    \nonumber
       &&=  \sum_{tuvw}\sum_{ijkl} \Phi_{tuvw} a_{ijklpqrs} \delta_{it}\delta_{ju}\delta_{kv}\delta_{lw} \\
       &&=  \sum_{tuvw} \Phi_{tuvw} a_{tuvwpqrs}. \label{tildestar2}
   \end{eqnarray}
So, using Eq.~(\ref{tildestar1}) and Eq.~(\ref{tildestar2}) we have 
\begin{eqnarray}
    \inner{\Theta^{*}( \mathcal{E}^{H_3 \rightarrow H_4}_{pqrs}), \Phi}=  \inner{ \Tilde{\Theta}( \mathcal{E}^{H_3 \rightarrow H_4}_{pqrs}), \Phi}.
\end{eqnarray}
As $\Phi$ is arbitrary, we conclude that $\Theta^{*}=\Tilde{\Theta}$.
\end{prof}
\begin{proposition}
  For a super-map $\Theta$, we always have $\Lambda_{\Theta}= \Lambda^{'}_{\Theta^{*}}$, where if $\Lambda= \sum_{i}\Psi_{i} \otimes \Phi_{i}$ then by $\Lambda^{'}$ we denote $\Lambda^{'}=\sum_{i}\Phi_{i} \otimes \Psi_{i}$. \label{prop5}
\end{proposition}
\begin{prof} The Choi-type representation of $\Theta^{*}$ is given by
   $ \Lambda_{\Theta^{*}}=\sum_{ijkl}  \mathcal{E}^{H_3 \rightarrow H_4}_{ijkl} \otimes \Theta^{*}( \mathcal{E}^{H_3 \rightarrow H_4}_{ijkl})$.
 Now we have
     $ \Lambda^{'}_{\Theta^{*}}=\sum_{ijkl} \Theta^{*}( \mathcal{E}^{H_3 \rightarrow H_4}_{ijkl}) \otimes \mathcal{E}^{H_3 \rightarrow H_4}_{ijkl}
      = \sum_{ijkl} \Tilde{\Theta}( \mathcal{E}^{H_3 \rightarrow H_4}_{ijkl}) \otimes \mathcal{E}^{H_3 \rightarrow H_4}_{ijkl}
      = \Lambda_{\Theta}$, where we have used Proposition(\ref{tilda=star}) in the second equality and Eq.~(\ref{tilde-choi-type}) in the last equality.
\end{prof}
\begin{proposition}
The bi-linear pairing between two super-map is equal to the bi-linear pairing between their adjoints i.e., $\inner{\Theta^{*}_1,\Theta^{*}_2}=\inner{\Theta_1,\Theta_2}$. \label{prop6}
\end{proposition}
\begin{prof} From the definition of bi-linear pairing we have $ \inner{\Theta^{*}_1,\Theta^{*}_2}= \inner{\Lambda_{\Theta^{*}_1},\Lambda_{\Theta^{*}_2}} = \inner{\Lambda^{'}_{\Theta^{*}_1}\Lambda^{'}_{\Theta^{*}_2}} = \inner{\Lambda_{\Theta_1}\Lambda_{\Theta_2}} = \inner{\Theta_1,\Theta_2}$,
    where we have used proposition (\ref{prop5}) in the third equality.
\end{prof}
\par Now we proceed towards our main theorems. Let us denote by $\mathcal{M}:= \sum_{ijkl}  \mathcal{E}^{H_1 \rightarrow H_2}_{ijkl} \otimes \mathcal{E}^{H_1 \rightarrow H_2}_{ijkl} $. Expanding $\mathcal{G} \in L(B(H_1),B(H_2)) \otimes L(B(H_1),B(H_2))$ in basis it can easily be shown that there exists a unique $\Theta^{\mathcal{G}} \in L[L(B(H_1), B(H_2))]$ such that $\mathcal{G}=(id \otimes \Theta^{\mathcal{G}}) \mathcal{M}$. We define $\Lambda^{\mathcal{G}}_{\Theta}=(id \otimes \Theta)\mathcal{G}= (id \otimes \Theta\circ \Theta^{\mathcal{G}}) \mathcal{M}$.
\\
So the statement:``$\Theta$ is completely CP-preserving if and only if $\Lambda^{\mathcal{G}}_{\Theta}$ is CP"  holds if and only  if 
\begin{eqnarray}
    \Theta \in \mathbb{CCPP}[(1,2) \rightarrow (3,4)] \iff \Theta\circ \Theta^{\mathcal{G}} \in \mathbb{CCPP} [(1,2) \rightarrow (3,4)], \label{ccpp1}
\end{eqnarray}
where $\mathbb{CCPP} [(1,2) \rightarrow (3,4)]$ represents the convex cone of completely CP-preserving maps in $L\left[L(B(H_1),B(H_2)), L(B(H_3),B(H_4))\right]$.
\\
Now $\Theta\circ \Theta^{\mathcal{G}} \in \mathbb{CCPP} [(1,2) \rightarrow (3,4)]$ holds if and only if $\forall \Delta \in \mathbb{CCPP} [(1,2) \rightarrow (3,4)]$
\begin{eqnarray}
\nonumber
&&  \inner{\Theta \circ \Theta^{\mathcal{G}}, \Delta} \geq 0 \\
\nonumber    
 &&  \implies { \inner{\Lambda_{\Theta \circ \Theta^{\mathcal{G}}}, \Lambda_{\Delta}} \geq 0} \\
\nonumber
&& \implies { \sum_{ijkl} \sum_{pqrs} \inner{\mathcal{E}^{H_1 \rightarrow H_2}_{ijkl} \otimes (\Theta \circ \Theta^{\mathcal{G}})(\mathcal{E}^{H_1 \rightarrow H_2}_{ijkl}), \mathcal{E}^{H_1 \rightarrow H_2}_{pqrs} \otimes \Delta(\mathcal{E}^{H_1 \rightarrow H_2}_{ijkl})} \geq 0 }\\
\nonumber
&& \implies { \sum_{ijkl} \sum_{pqrs} \inner{\mathcal{E}^{H_1 \rightarrow H_2}_{ijkl},\mathcal{E}^{H_1 \rightarrow H_2}_{pqrs}} \inner{ (\Theta \circ \Theta^{\mathcal{G}})(\mathcal{E}^{H_1 \rightarrow H_2}_{ijkl}), \Delta(\mathcal{E}^{H_1 \rightarrow H_2}_{ijkl})} \geq 0 } \label{uprop3} \\
\nonumber
&& \implies { \sum_{ijkl} \sum_{pqrs} \inner{\mathcal{E}^{H_1 \rightarrow H_2}_{ijkl},\mathcal{E}^{H_1 \rightarrow H_2}_{pqrs}} \inner{  \Theta^{\mathcal{G}}(\mathcal{E}^{H_1 \rightarrow H_2}_{ijkl}), \Theta^{*} \circ \Delta(\mathcal{E}^{H_1 \rightarrow H_2}_{ijkl})} \geq 0 } \\
\nonumber
 && \implies { \sum_{ijkl} \sum_{pqrs} \inner{\mathcal{E}^{H_1 \rightarrow H_2}_{ijkl} \otimes  \Theta^{\mathcal{G}}(\mathcal{E}^{H_1 \rightarrow H_2}_{ijkl}), \mathcal{E}^{H_1 \rightarrow H_2}_{pqrs} \otimes \Theta^{*} \circ \Delta(\mathcal{E}^{H_1 \rightarrow H_2}_{ijkl})} \geq 0 } \label{u1prop3}
 \end{eqnarray}
 where we have used Proposition~(\ref{prop3}) in the third and the last implications. From the last implication we have $\inner{\Lambda_{\Theta^{\mathcal{G}}}, \Lambda_{\Theta^{*} \circ\Delta}} \geq 0$, which implies ${\inner{\Theta^{\mathcal{G}}, \Theta^{*} \circ \Delta} \geq 0 }$. Using Proposition~(\ref{prop6}) and proposition~(\ref{prop2}) it can easily be seen that $\inner{\Theta^{\mathcal{G}}, \Theta^{*} \circ \Delta} \geq 0 $ implies ${ \inner{\Theta^{\mathcal{G} *},\Delta^{*} \circ \Theta} \geq 0}$ which implies ${ \inner{\Delta^{*} \circ \Theta, \Theta^{\mathcal{G} *}} \geq 0 }$ which further implies ${ \inner{\Theta, \Delta \circ \Theta^{\mathcal{G} *}} \geq 0}$.
Now the inequality $\inner{\Theta, \Delta \circ \Theta^{\mathcal{G} *}} \geq 0$ implies that $\Theta \in (\mathbb{CCPP} \circ \{ \Theta^{\mathcal{G}*}\})^{\circ}$. 
\par So from (\ref{ccpp1}) we have that the CCPP vs CP correspondence holds if and only if
\begin{center}
    $\mathbb{CCPP}[(1,2) \rightarrow (3,4)]=(\mathbb{CCPP}[(1,2) \rightarrow (3,4)] \circ \{ \Theta^{\mathcal{G}*}\})^{\circ}$.
\end{center}
From proposition(\ref{prop1}), we have $\mathbb{CCPP}^{\circ}=\mathbb{CCPP}$. Now $\mathbb{CCPP} \circ \{ \Theta^{\mathcal{G}*}\}$ is convex and topologically closed which implies that $(\mathbb{CCPP} \circ \{ \Theta^{\mathcal{G}*}\})^{\circ \circ}=\mathbb{CCPP} \circ \{ \Theta^{\mathcal{G}*}\}$. So, we finally have that the CCPP vs CP correspondence holds if and only if
\begin{center}
    $\mathbb{CCPP}[(1,2) \rightarrow (3,4)]=\mathbb{CCPP}[(1,2) \rightarrow (3,4)] \circ \{ \Theta^{\mathcal{G}*}\}$\\
    if and only if \\
   $ \mathbb{CCPP}[(3,4) \rightarrow (1,2)]=\{ \Theta^{\mathcal{G}}\} \circ \mathbb{CCPP}[(3,4) \rightarrow (1,2)]$.
\end{center}
Note that the last equation is true for any choices of Hilbert spaces $H_1, H_2, H_3$ and $H_4$. So when we take $H_3$ and $H_4$ same as $H_1$ and $H_2$ respectively then we have the identity super-map $id \in \mathbb{CCPP}[(1,2) \rightarrow (1,2)] $. So we have both $\Theta^{\mathcal{G}}$ and ${\Theta^{\mathcal{G}}}^{-1}$ completely CP-preserving which implies that $\Theta^{\mathcal{G}}$ is a complete order isomorphism.
\begin{theorem}
    The CCPP vs CP correspondence holds for $\mathcal{G} \in L(B(H_1),B(H_2)) \otimes L(B(H_1),B(H_2))$ if and only if $\Theta^{\mathcal{G}}$ is a complete order isomorphism.
\end{theorem}
Now we will provide the necessary and sufficient condition on a basis $\{ \mathcal{F}^{H_1 \rightarrow H_2}_{\beta} \}$ such that that CP vs CCPP correspondence holds.
\\
We denote by $\mathcal{N}:= \sum_{\beta}  \mathcal{F}^{H_1 \rightarrow H_2}_{\beta} \otimes \mathcal{F}^{H_1 \rightarrow H_2}_{\beta} $ and $\Lambda^{\mathcal{N}}_{\Theta} := \sum_{\beta}  \mathcal{F}^{H_1 \rightarrow H_2}_{\beta} \otimes \Theta(\mathcal{F}^{H_1 \rightarrow H_2}_{\beta}) $. To simplify the notation we rename the basis $\{ \mathcal{E}^{H_1 \rightarrow H_2}_{ijkl}  \}$ by $\{ \mathcal{E}_{\alpha}\}$. Let $V$ denotes the transformation that transforms the basis $\{ \mathcal{E}_{\alpha}\}$ into the basis $\{ \mathcal{F}_{\beta} \}$, i.e, $\mathcal{F}_{\alpha} =V(\mathcal{E}_{\alpha})=\sum_{\beta} V_{\beta \alpha} \mathcal{E}_{\beta}$. Now we have
\begin{eqnarray}
\nonumber
    \mathcal{N} &&=\sum_{\beta,\gamma}  \mathcal{E}_{\beta} \otimes \sum_{\alpha} V_{\beta \alpha} V_{\gamma \alpha}\mathcal{E}_{\gamma}\\
\nonumber    
    &&= \sum_{\beta,\gamma}  \mathcal{E}_{\beta} \otimes \sum_{\alpha} V_{\beta \alpha} V^{t}_{\alpha \gamma }\mathcal{E}_{\gamma}\\
\nonumber    
    &&= \sum_{\beta,\gamma}  \mathcal{E}_{\beta} \otimes (V V^{t})_{\beta \gamma }\mathcal{E}_{\gamma}\\
\nonumber    
    &&= \sum_{\beta}  \mathcal{E}_{\beta} \otimes \sum_{\gamma}(V V^{t})_{\gamma \beta }\mathcal{E}_{\gamma}\\
\nonumber    
    &&= \sum_{\beta}  \mathcal{E}_{\beta} \otimes(V V^{t})\mathcal{E}_{\beta}\\
\nonumber    
    &&= (id \otimes V V^{t}) \sum_{\beta}  \mathcal{E}_{\beta} \otimes \mathcal{E}_{\beta}.
\end{eqnarray}
Denoting by $\Theta^{\mathcal{N}}$=$VV^{t} \in L[L(B(H_1),B(H_2))]$,  we have
\begin{eqnarray}
    \mathcal{N}=(id \otimes \Theta^{\mathcal{N}}) \mathcal{M},
\end{eqnarray}
which implies 
\begin{eqnarray}
    \Lambda^{\mathcal{N}}_{\Theta}= (id \otimes \Theta\circ \Theta^{\mathcal{N}}) \mathcal{M},
\end{eqnarray}
and this leads to the following theorem
\begin{theorem}
    The CCPP vs CP correspondence holds for a basis $\{ \mathcal{F}^{H_1 \rightarrow H_2}_{\beta} \}$ if and only if $\Theta^{\mathcal{N}}$ is a complete order isomorphism.
\end{theorem}
\subsection{Structure of complete order isomorphism}
A super-map $\Theta :L(B(H_1),B(H_2)) \rightarrow L(B(H_3),B(H_4))$ induces a map $T: B(H_1)\otimes B(H_2) \rightarrow B(H_3)\otimes B(H_4) $ at the level of Choi matrices and is given by the following:
\begin{eqnarray}
    T(x)= C_{\Theta(\Gamma_{x})}.
\end{eqnarray}
Following \cite{Chiribella}, we will call this map the ``representing map" for the super-map $\Theta$.
Now a given map $T:B(H_1)\otimes B(H_2) \rightarrow B(H_3)\otimes B(H_4)$ induces a super-map $\Theta : L(B(H_1),B(H_2)) \rightarrow L(B(H_3),B(H_4))$ as follows:
\begin{eqnarray}
    \Theta(\Phi)(A)=Tr_{3}[(A^{t} \otimes \mathbb{I}) T(C_{\Phi})]. \label{inv_rep}
\end{eqnarray}
\begin{proposition}
The map $\Theta \rightarrow T$ is an algebra isomorphism between the spaces \\$L[L(B(H_1),B(H_2)), L(B(H_3),B(H_4))]$ and $L[B(H_1)\otimes B(H_2),B(H_3)\otimes B(H_4)]$.
\end{proposition}
\begin{prof}
Assume that $\Theta_1 \neq \Theta_2$. This implies that $ \Theta_1(\Psi) \neq \Theta_2(\Psi)$ for all $\Psi \in L(B(H_1),B(H_2))$. Using the Choi-Jamio{\l}kowski isomorphism we have $C_{\Theta_1(\Psi)} \neq C_{\Theta_2(\Psi)}$. Now from the definition of representing map we can write the last relation as $T_{1}(C_{\Psi})\neq T_{2}(C_{\Psi})$ which implies $T_1 \neq T_2$. So $\Theta \rightarrow T$ is injective. It follows from Eq.~(\ref{inv_rep}) that this map is surjective as well. So the map $\Theta \leftrightarrow T$ is one to one correspondence.
\\
\emph{Linearity}: Let $\Theta_1 \leftrightarrow T_1$, $\Theta_2 \leftrightarrow T_2$ and $\alpha \Theta_1 + \beta \Theta_2\leftrightarrow T$, where $\alpha$ and $\beta$ are arbitrary complex numbers. Now $T(x)=T(C_{\Gamma_{x}})=C_{\alpha \Theta_1 (\Gamma_{x}) + \beta \Theta_2 (\Gamma_{x})} $. Using Choi-Jamio{\l}kowski isomorphism and the definition of the representing map, we have $C_{\alpha \Theta_1 (\Gamma_{x}) + \beta \Theta_2 (\Gamma_{x})}= \alpha T_{1}(C_{\Gamma_{x}}) + \beta T_{2}(C_{\Gamma_{x}})=(\alpha T_{1} + \beta T_{2})x$. As $x$ is arbitrary we conclude that $T=\alpha T_1 + \beta T_2$.\\
\emph{Product preserving}: Let $T \leftrightarrow \Theta_1 \circ \Theta_2$. Now $ T(x)=T(C_{\Gamma_{x}})=C_{\Theta_{1} (\Theta_{2} (\Gamma_{x}))}$. From the definition of the representing map we have $C_{\Theta_{1} (\Theta_{2} (\Gamma_{x}))}=T_{1} (C_{\Theta_{2} (\Gamma_{x})})= T_{1} (T_{2}(C_{\Gamma_{x}}))= (T_{1} \circ T_{2}) x$, which implies $T=T_{1} \circ T_{2}$.\\
So we conclude that the map $\Theta \leftrightarrow T$ is an algebra isomorphism.
\end{prof}
\begin{theorem} \cite{Chiribella,Gour}
  A super-map $\Theta$ is completely CP-preserving if and only if its representing map $T$ is CP.
\end{theorem}
\begin{proposition}
   A super-map $\Theta$ is complete order isomorphism if and only if its representing map $T$ is complete order isomorphism.
\end{proposition} 
\begin{prof}
  A super-map $\Theta$ is complete order isomorphism if and only if both $\Theta$ and $\Theta^{-1}$ are completely CP-preserving if and only if both $T$ and $T^{-1}$ are CP if and only if $T$ is complete order isomorphism.
\end{prof}
\\
In the following theorem we give the explicit mathematical form of a complete order isomorphism. 
\begin{theorem}
   The super-map $\Theta :L(B(H_1),B(H_2)) \rightarrow L(B(H_3),B(H_4))$ is a complete order isomorphism if and only if there exists invertible $K: H_1 \otimes H_2 \rightarrow H_3 \otimes H_4$ such that
    \begin{eqnarray}
        \Theta(\Phi)(A)=Tr_{3}[(A^{t} \otimes \mathbb{I}) KC_{\Phi}K^{\dagger}].
    \end{eqnarray}
\end{theorem}
\begin{prof}
  From (\ref{KPCOI}) we see that the representing map $T: B(H_1)\otimes B(H_2) \rightarrow B(H_3)\otimes B(H_4)$ is a complete complete order isomorphism if and only if there exists invertible $K: H_1 \otimes H_2 \rightarrow H_3 \otimes H_4$ such that $T(X)= K X K^{\dagger}$. Now we use Eq.~(\ref{inv_rep}) to arrive at the result.
\end{prof}
\section{Conclusion}\label{s4}
The Choi-Jamio{\l}kowski-Kraus-Sudharshan isomorphism states that a map between two matrix algebras is completely positive if and only if the so-called Choi matrix is positive. This complete positivity vs positivity correspondence depends on the basis we choose to define the Choi matrix. The work of Paulsen and Shultz, and later the work of Kye, provide the necessary and sufficient condition on a basis for the correspondence to hold. A similar  correspondence holds for super-maps and maps as well. It states that a super-map is completely CP-preserving if and only if its Choi-type representation is CP. And here again, the correspondence depends on the choice of basis. In this work, we have presented the necessary and sufficient condition on a basis such that the completely CP-preserving vs CP correspondence holds true. 
\\
\\
\emph{Acknowledgments.} We are thankful to S.H. Kye for useful discussions. US acknowledges partial support from the Interdisciplinary Cyber-Physical Systems (ICPS) program of the Department of Science and Technology (DST), Government of India, Grant No.~DST/ICPS/QuST/Theme-3/2019/120. RS acknowledges financial support  from DST Project No. {\sf DST/ICPS/QuST/Theme-2/2019/General Project  Q-90}.
\bibliographystyle{apsrev4-1}
\bibliography{referencefile}

\end{document}